\newcommand{\PLA}[3]{Phys.\ Lett.\ A\ {\bf #1},\ #2 (#3)}

\newcommand{\PRL}[3]{Phys.\ Rev.\ Lett.\ {\bf #1},\ #2 (#3)}

\newcommand{\OL}[3]{Opt.\ Lett.\ {\bf #1},\ #2 (#3)}
\newcommand{\OC}[3]{Opt.\ Commun.\ {\bf #1},\ #2 (#3)}

\newcommand{\PRE}[3]{Phys.\ Rev.\ E\ {\bf #1},\ #2 (#3)}

\newcommand{\JETP}[3]{JETP.\ Lett.\ {\bf #1},\ #2 (#3)}

\newcommand{\JOSAB}[3]{J.\ Opt.\ Soc.\ Am.\ B\ {\bf #1},\ #2 (#3)}

\newcommand\md{\mathrm{d}}

\newcommand{\SCR}{Schr\"odinger~}

\newcommand{\sech}{\mathrm{sech}}
\newcommand{\diracslash}[1]{#1\llap{/\kern2pt}}

\newcommand{\be}{\begin{equation}}
\newcommand{\ee}{\end{equation}}
\newcommand{\bea}{\begin{eqnarray}}
\newcommand{\eea}{\end{eqnarray}}
\newcommand{\ba}[1]{\begin{array}{#1}}
\newcommand{\ea}{\end{array}}

\documentclass[twocolumn,superscriptaddress,secnumarabic,amssymb,nobibnotes,nofootinbib,aps,prl,showpacs]{revtex4}
\usepackage{graphicx}
\usepackage{color}
\begin{document}
\setlength{\topmargin}{-0.05in}
\title{Controlling pulse propagation in optical fibers through nonlinearity and dispersion management}

\author{Rajneesh Atre}\thanks{atre@prl.res.in}
\affiliation{Physical Research Laboratory, Navrangpura, Ahmedabad
380 009, India} \affiliation{School of Physics, University of
Hyderabad, Hyderabad-500 046, India}

\author{Prasanta K. Panigrahi}\thanks{prasanta@prl.res.in}
\affiliation{Physical Research Laboratory, Navrangpura, Ahmedabad
380 009, India}

\date{\today}

\begin{abstract}  In  case  of  the  nonlinear  \SCR equation with
designed  group  velocity  dispersion,  variable  nonlinearity  and
gain/loss;  we  analytically  demonstrate  the  phenomenon  of chirp
reversal  crucial for pulse reproduction. Two different scenarios are
exhibited,  where  the  pulses  experience  identical  dispersion
profiles,  but  show  entirely different propagation behavior. Exact
expressions for dynamical quasi-solitons and soliton bound-states relevant
for fiber communication are also exhibited.
\end{abstract}


\maketitle  Nonlinear \SCR equation (NLSE) is known to govern the
pulse  dynamics  in  nonlinear  optical fibers \cite{GPA}. In recent
years,  the  study  of  nonlinear fiber optics, dealing with optical
solitons,  has  attracted  considerable  attention  since  it has an
important  role  in  the  development of several technologies of the
21st  century \cite{HasegawaB}. NLSE with distributed coefficients
such  as,  group velocity dispersion (GVD), distributed nonlinearity
and  gain/loss,  is  being studied extensively in order to determine
the  effect  of  various  distributed  parameters on the pulse
profile.

In  the realistic situation in a fiber, there arises non-uniformity
due to variation in the lattice parameters of the fiber medium, as a
result  of  which the distance between two neighboring atoms is not
constant  throughout  the  fiber.  It  may  also  arise  due to the
variation of the fiber  geometry  e.g.,  diameter fluctuation. These
non-uniformities influence  effects  such  as, loss (or gain), phase
modulation,  etc,  which  can  be  modeled  by making corresponding
parameters   space  dependent.  From  a  practical  point  of  view,
tailoring  of various fiber parameters may lead to effective control
of  the pulse. This has been one of the prime motivation of a number
of   authors  to  analyze  NLSE  in  a  distributed  scenario.

Dispersion  management  (DM) has emerged as an important technology to
control  and  manipulate  the  light  pulses  in  optical fibers
\cite{HasegawaB,Ablowitz}.  Pulse  compression has been demonstrated with
appropriately designed GVD and nonlinearity in the presence of chirping
\cite{Moores,Kruglov,PKP}, as also through soliton effects
\cite{Mollenauer}.   Adiabatic   soliton  compression,  through  the
decrease  of dispersion along the length of the fiber has been shown to
provide  good  pulse  quality  \cite{Dianov}. The possibility of
amplification   of   soliton   pulses  using  a  rapidly  increasing
distributed  amplification  with  scale  lengths  comparable  to the
characteristic  dispersion length has been reported \cite{Lisak}. It has
been  numerically shown that, in the case where the gain due to the
nonlinearity  and  the  linear  dispersion  balance each other,
equilibrium solitons are formed \cite{Malomed1}. Serkin and Hasegawa have
formulated the effect of varying dispersion and other parameters on   the
soliton  dynamics  and  have  explained  the  concept  of amplification of
soliton \cite{Serkin}.

The  formal structure of the Lax pair for the deformed NLSE has been
studied  \cite{Brustev,SerkinPRL}.  In  a significant result, it has been
numerically  shown  that in an appropriately designed dispersion profile
chirped pulses can be retrieved through chirp reversal at a calculated
location  in  the  fiber  \cite{Kumar}. The advantage of pre-chirping  of
the  input pulse in overcoming soliton interaction and dispersive-wave
generation has been noted earlier \cite{Gabitov}.

In the present Letter, we demonstrate analytically the phenomenon of
chirp reversal of quasi-solitons with a designed dispersion profile.
Very  interestingly  we find two possibilities of chirp reversal
for  which  the  dispersion  profiles  are  identical. However, they
exhibit  entirely  different  propagation  behavior. In one case the
motion   is   sinusoidal   and  in  the  other  it  shows  pulse
acceleration.  The  procedure  to  control  pulse dynamics is also
pointed  out. Exact expressions for dynamical quasi-solitons and
soliton bound-states relevant for fiber communication are exhibited.

It  is  worth  emphasizing that, exact solutions have played crucial
role  in  demonstrating different pulse  shaping techniques.  The
soliton  solutions of NLSE or modifications of the same  has  come
in  handy in studying these mechanisms. In the same light,  finding
exact  solutions  for  general types of distributed scenarios  will
illustrate  the  subtle  effects  and  interplay of various
parameters  on  formation and propagation dynamics of light pulses.

We  develop  a  methodology  to  obtain  self-similar  solitary wave
solutions  of generalized NLSE model with varying nonlinearity, GVD,
gain/loss and a confining oscillator which can be further modulated.
One  or few of these parameters can be switched off depending on the
situation  at  hand.  It is shown that, this equation decouples into
elliptic  function  equation  and  a  \SCR  eigenvalue problem. This
allows one to analytically treat a variety of distributed scenarios,
a  few  of  which  we  explicate  in  the  text.  In  the context of
Bose-Einstein  condensates  the  procedure  to  deal  with  variable
coefficient NLSE in the absence of GVD has been carried out recently
by the present authors \cite{Atre}. GVD leads to a fundamentally new
control parameter in the present case dealing with optical fibers.
For example, a subtle  interplay  of  GVD  and  nonlinearity  leads
to a soliton  bound-state  as  will  be  seen  below. The effect of
GVD, alternating  between  normal  and  anomalous dispersion,  on
pulse profile is also discussed.

For the purpose of analytic demonstration of chirp reversal we start
with  a  NLSE  model  with  variable GVD, nonlinearity and loss/gain
\cite{Kumar}:
\bea\label{GNLSE}
i{\partial_{z}}q(z,t)+\frac{d_{e}(z)}{2}\partial_{tt}q(z,t)+\gamma(z)|q|^{2}q(z,t)+ig(z)q(z,t)=0.
\eea

With $q(z,t)=a(z)u(z,t)$ and $a(z)=\exp[-\int_{0}^{z}dz'g(z')]$, one
obtains,

\bea
i{\partial_{z^{\prime}}}u(z^{\prime},t)+\frac{d_{e}(z')}{2}\partial_{tt}u(z^{\prime},t)+
\tilde{\gamma}(z')|u|^{2}u(z^{\prime},t)=0, \eea

where $z^{\prime}=\int_{0}^{z}dz^{\prime\prime} a^{2}(z^{\prime\prime}).$

Keeping in mind, pre-chirping and self-similar nature of the pulse
we make use of the following ansatz,

\be u(z',t)=\sqrt{p(z')}~\nu\left[ p(z')t,z'\right]\exp\left[i
C(z')t^{2}\right],\ee where, $p$ and $C$ are real functions of $z'$.

Defining $\tau= p(z')t, $ for preserving space-time identity one
obtains \bea i\left[\frac{\partial\nu}{\partial
z'}+K_{0}\tau\frac{\partial \nu}{\partial \tau}
\right]&+&\frac{d_{e}p^{2}}{2}\frac{\partial^{2}\nu(z^{\prime},\tau)}{\partial\tau^{2}}+
\tilde{\gamma}(z')p|\nu|^{2}\nu(z^{\prime},\tau) \nonumber\\
&-&\frac{K_{1}\tau^{2}}{2}\nu+\frac{iK_{0}\nu}{2}=0, \eea where
\be\label{KEq}
K_{0}=\frac{p_{z'}+Cd_{e}p}{p},{\mathrm{and}}~K_{1}=\frac{C_{z'}+C^{2}d_{e}}{p^{3}}.\ee
We now tailor the dispersion profile with $K_{0}=0$ and
$K_{1}={\mathrm{const}}$:

\bea i\frac{\partial\nu}{\partial
z'}+\frac{d_{e}p^{2}}{2}\frac{\partial^{2}\nu(z^{\prime},\tau)}{\partial\tau^{2}}+
\tilde{\gamma}(z')p|\nu|^{2}\nu(z^{\prime},\tau)=\frac{K_{1}\tau^{2}}{2}\nu.\eea

In order to map the above equation to one with constant anomalous
dispersion we assume the constraint $d_{e}p=1$ to obtain,

\be\label{TNLSE} i\frac{\partial\nu}{\partial
z''}+\frac{1}{2}\frac{\partial^{2}\nu(z^{\prime\prime},\tau)}{\partial\tau^{2}}+
\tilde{\gamma}(z'')|\nu|^{2}\nu(z^{\prime},\tau)=\frac{K_{1}\tau^{2}}{2}\nu,
\ee where $z''=\int_{0}^{z'}p(s)ds$.

The  above equation has been numerically investigated, where a
chirp-reversal  was  observed  for  quasi-soliton  having  a profile
intermediate   to   a   Gaussian   and   the fundamental   NLSE soliton
\cite{Kumar}.  These  are stationary solutions obeying NLSE with an
additional oscillator term, which explains the above profile. The exact
solutions  of Eq. (\ref{TNLSE}) can be obtained, following the formalism
developed in Ref. \cite{Atre}:

\be
\nu(z^{\prime\prime},\tau)=\sqrt{A(z^{\prime\prime})}F[A(z^{\prime\prime})\{\tau-\Lambda(z^{\prime\prime})\}
]e^{i\Phi(z^{\prime\prime},\tau)}, \ee where \be
\Phi(z^{\prime\prime},\tau)=a_{1}(z^{\prime\prime})+b(z^{\prime\prime})\tau-
\frac{1}{2}c(z^{\prime\prime})\tau^2.\ee

Here,
$a_{1}(z^{\prime\prime})=a_{1_{0}}-\frac{\lambda-1}{2}\int_{0}^{z^{\prime\prime}}A^{2}(z)dz$,
$\Lambda(z^{\prime\prime})=\int_{0}^{z^{\prime\prime}}v(z)dz$, which
satisfies the following equation: \be \frac{\md \Lambda}{\md
z^{\prime\prime}}
+c(z^{\prime\prime})\Lambda(z^{\prime\prime})=b(z^{\prime\prime}),\ee
with the general solution,

\be\label{sol-centre} \Lambda(z^{\prime\prime})= \{e^{\int
c(z^{\prime\prime}) dz^{\prime\prime}}\}^{-1}\times \left[ \int
dz^{\prime\prime}\left\{b(z^{\prime\prime})e^{\int
c(z^{\prime\prime}) dz^{\prime\prime}}\right\}\right]. \ee

The parameter $c({z^{\prime\prime}})$ obeys Riccati equation:
\be\label{Rica}
c_{z^{\prime\prime}}-c^{2}(z^{\prime\prime})=K_{1},\ee which can be
exactly mapped to linear \SCR eigenvalue problem. We also find the
following consistency conditions:

\begin{equation}\label{consi}
\left.
\begin{array}{c}
{\tilde{\gamma}}(z^{\prime})={\tilde{\gamma}}_{0}A(z^{\prime})/A_{0},~~b(z^{\prime\prime})=A(z^{\prime\prime}) \\
A(z^{\prime\prime})=A_{0}\exp\left\{
\int_{0}^{z^{\prime\prime}}c(z)dz\right\},~~A_{0}>0.
\end{array}
\right\}
\end{equation}

$F$ obeys elliptic function equation in new variable
$T=A(z^{\prime\prime})\{\tau-\Lambda(z^{\prime\prime})\}$:

\be \label{elliptic} F''(T)-\lambda F(T)+2\kappa F^{3}(T)=0,
~{\mathrm{where}}~ \kappa=-\frac{\gamma_{0}}{A_0}. \ee

The twelve Jacobian elliptic functions satisfy above equation. These
functions interpolate between the trigonometric and hyperbolic functions
in the limiting cases \cite{Hancock}. Bright soliton solutions of the type
$\nu(z^{\prime\prime},\tau)=\sqrt{A(z^{\prime\prime})}~\sech[T/T_{0}]e^{i\Phi(z^{\prime\prime},\tau)}$
exist for $\kappa>0$, where $T_{0}^{2}=-A_{0}/\gamma_{0}$ and
$\lambda=1/2T_{0}^{2}$, similarly kink-type dark solitons exist for
$\gamma_{0}>0$. We further note that, with normal dispersion one can
obtain dark solitons for $\gamma_{0}<0$. It needs to be emphasized that,
in the present approach the oscillator term leads to a dynamical chirp and
modulates the pulse profile.  However, the pulse retains its fundamental
NLSE soliton character in the scaled variable $z^{\prime\prime}$.

Below, we examine the formation of bright quasi-soliton like excitations,
exhibiting chirp reversal phenomenon. This is accomplished by appropriate
tailoring of GVD and pre-chirping of the launching pulse. Combining Eq.
(\ref{KEq}) and constraint $d_{e}p=1$, yield the following expressions for
the tailored dispersion and the chirping parameter,

\be\label{TailDisp}
C=\frac{d_{e}^{\prime}}{d_{e}^{2}},~~\mathrm{and}~~d_{e}\frac{\md^{2}d_{e}}{\md
z^{\prime 2}}-\left( \frac{\md d_{e}}{\md
z^{\prime}}\right)^{2}=K_{1}.\ee

It is interesting to notice that the choice of the constant $K_{1}$,
gives  rise  to two scenarios, having identical dispersion and chirp
profiles,   but  possessing  entirely  different  pulse  velocities.

We list below some explicit examples, depicting a variety of novel
control mechanism for pulse manipulation.

\noindent{\it{Soliton pulses exhibiting chirp-reversal.--}} Inspired from
the numerical investigations of Kumar and Hasegawa \cite{Kumar}, we first
consider in Eq. (\ref{TailDisp}), $K_{1}>0$, which  is equivalent  to a
regular oscillator potential in Eq. (\ref{TNLSE}). Dispersion profile in
this case reads,
\be\label{disp_pos} d_{e}(z^{\prime})=\cosh(\delta
z^{\prime})+\frac{C(0)}{\delta}\sinh(\delta
z^{\prime}),{\mathrm{with}}~\delta=\left[K_{1}+C^{2}(0) \right]^{1/2}.\ee

\begin{figure}
\includegraphics[width=2.5in]{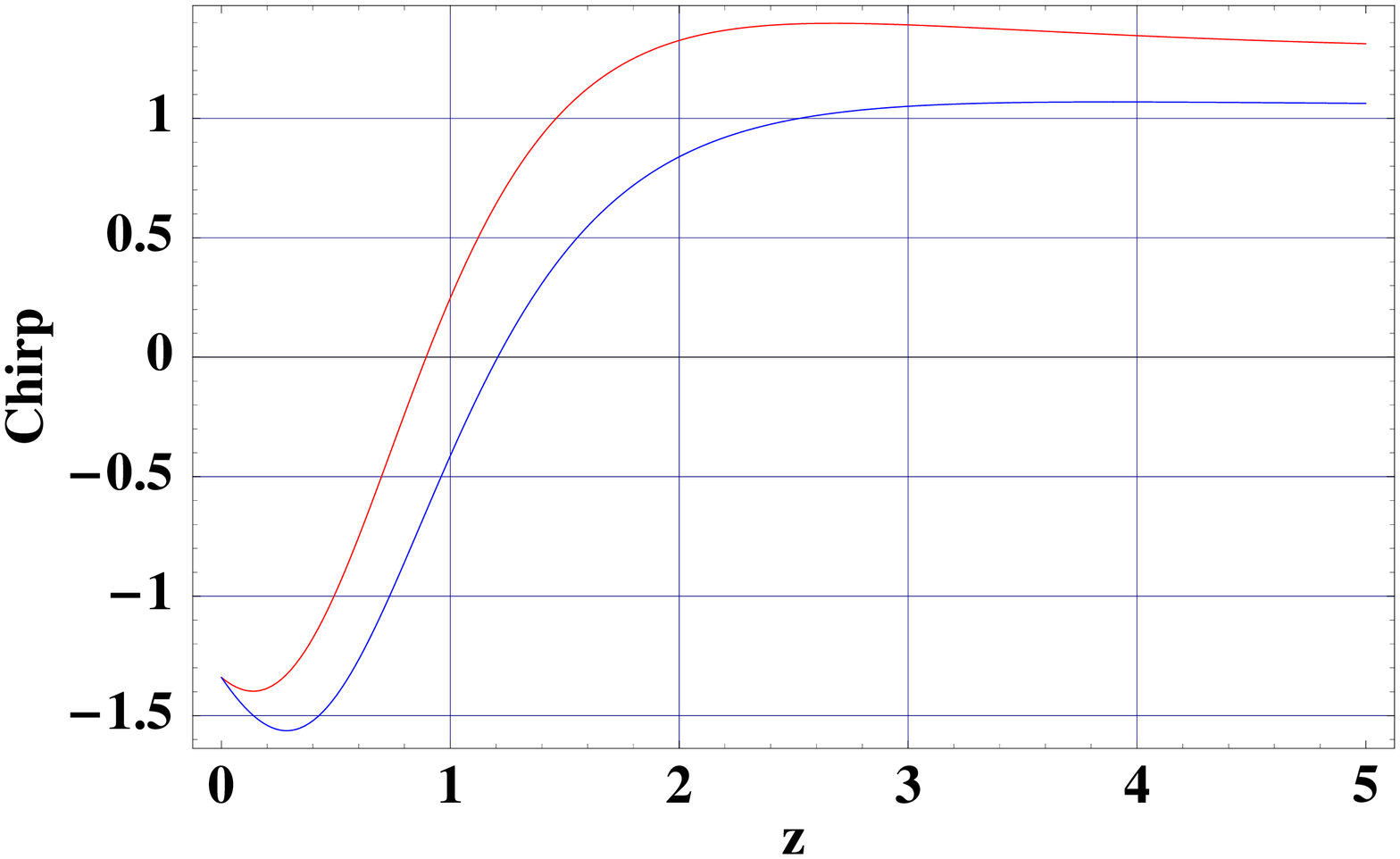}
\caption{The figure depicting variation of chirp parameters with $z$. The
red curve shows the launching chirp and the blue curve shows the same in
combination with the dynamical chirp.}\label{Chirp_reg}
\end{figure}

From above dispersion profile we compute launched chirp parameter
$C(z^{\prime})$ and plot it together with the dynamical chirp in Fig.
\ref{Chirp_reg}. From Eqs. (\ref{KEq}) and (\ref{disp_pos}) it is clear
that launched chirp profile changes  sign at
$z(z^{\prime}_{c})=(1/\delta)\tanh^{-1}[-C(0)/\delta].$

\begin{figure}
\includegraphics[width=2.5in]{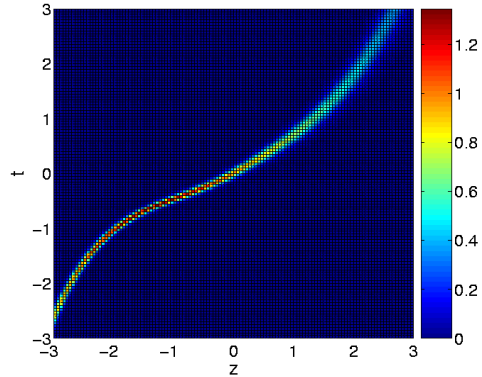}
\caption{Pulse propagation having sinusoidal velocity profile. The plot
shows broadening after chirp-reversal.}\label{sin_pulse}
\end{figure}

The expression for traveling quasi-soliton, propagating with
velocity $v(z^{\prime\prime})=A_{0}\cos\left(K_{0}
 z^{\prime\prime}\right),$ reads
\bea q(z,t)&=&a(z)\sqrt{p(z^{\prime})\sec(K_{1}z^{\prime\prime})}~
\sech\left[\sec(K_{1}z^{\prime\prime})\{\tau-\Lambda(z^{\prime\prime})
\}\right]\nonumber\\&\times&\exp\left\{{i\left[C(z^{\prime})t^{2}+\Phi(z^{\prime\prime},\tau)\right]}\right\}.
\eea
The presence of the dynamical chirp shifts the chirp-reversal location
slightly away from $z_{c}$. Just after chirp reversal, we notice that the
pulse seems to broaden, as is clearly seen in Fig. \ref{sin_pulse}. Hence,
the pulse needs to be retrieved at this point. With the help of a normal
dispersive element such as a grating the original pulse can be recovered
\cite{Belanger}.

In contrast to the above example, if we consider $K_{1}<0$, we
obtain $\delta=[K_{1}-C^{2}(0)]^{1/2}$, which again leads to the
same dispersion profile as that of the previous example, but with
the pulse velocity $v(z^{\prime\prime})=A_{0}\cosh\left(K_{0}
 z^{\prime\prime}\right).$ The expression for the soliton profile in
 this case is,

\bea q(z,t)&=&a(z)\sqrt{p(z^{\prime})\sech(K_{1}z^{\prime\prime})}
\sech\left[\sech(K_{1}z^{\prime\prime})\{\tau-\Lambda(z^{\prime\prime})
\}\right]\nonumber\\&\times&\exp\left\{{i\left[C(z^{\prime})t^{2}+\Phi(z^{\prime\prime},\tau)\right]}\right\}.
\eea

It is interesting to observe that, compared with the previous case
pulse broadening is significantly reduced for the same parameter
values.

\begin{figure}
\includegraphics[width=2.5in]{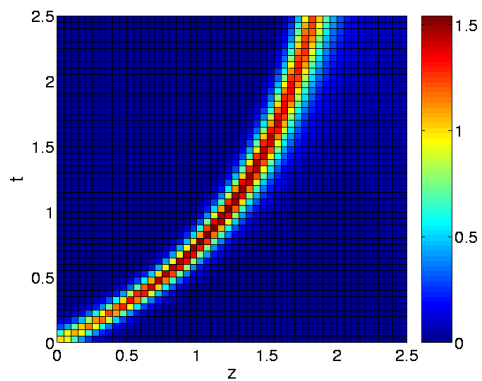}
\caption{Soliton pulse acceleration with parameter $K_{1}<0$
implying an expulsive oscillator scenario.}\label{Pulse_Expuls}
\end{figure}

\noindent{\it{Soliton bound-states.--}} Starting from Eq. (\ref{GNLSE})
without  tailoring  the  dispersion  profile, we proceed to obtain
self-similar solutions assuming the ansatz solution of the type:

\be\label{bst-ansatz}
q(z,t)=\sqrt{A(z)}F\left[A(z)\left\{\tau-\Gamma(z)
\right\}\right]\exp\left\{G(z)+i\Phi(z,t) \right\}.\ee

The parameters appearing in the Eq. (\ref{bst-ansatz}) can be
straightforwardly evaluated from the Eqns. (\ref{sol-centre}),
(\ref{Rica}), (\ref{consi}) and soliton profile can be obtained from Eq.
(\ref{elliptic}).

\begin{figure}
\includegraphics[width=2.5in]{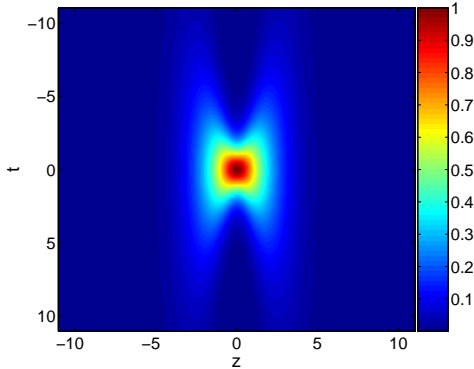}
\caption{Soliton bound-state in a medium with two dispersion
regimes.}\label{sptbst}
\end{figure}

\begin{figure}
\includegraphics[width=2.5in]{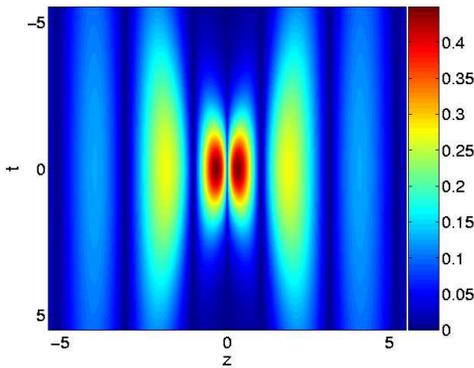}
\caption{Multiple soliton bound-state with oscillatory
gain/loss.}\label{Multbst}
\end{figure}

Below, we explicate some examples of spatial bound-states of solitons,
arising from interplay of GVD, nonlinearity and gain/loss. Below, Fig.
\ref{sptbst} depicts a  two-soliton bound-state. This arises in a medium,
where both anomalous and normal dispersion regimes are smoothly connected.
In the presence of periodic gain/loss one observes modulation in the
bound-state profile as is shown in Fig. \ref{Multbst}.

In conclusion, we have obtained exact soliton solutions exhibiting
chirp-reversal, while retaining their original profile, crucial for pulse
recovery in fiber optics. Two different soliton sectors differing in
propagation behavior, but with identical dispersion profiles, are
analytically exhibited. We have outlined a general formalism for obtaining
self-similar  solutions  of nonlinear \SCR equation, in the presence of
distributed  coefficients, from which earlier scenarios follow as special
cases \cite{Kruglov1}. It is shown that, this nonlinear system, involving
pulse propagation with group velocity dispersion (GVD), variable
nonlinearity, variable gain, exactly decouples into elliptic function
equation and a \SCR eigenvalue problem. This opens up a number  of
possibilities  to take  into account a wide class of distributed
scenarios, in close conformity with the experimentally achievable
situations. This incorporates a number of special cases dealt earlier,  in
the context  of  pulse compression.  We find that, apart  from
compression, one can achieve  control over the pulse velocity, pulse
profile, through interplay  of  group velocity dispersion, nonlinearity,
gain/loss. Formation of soliton bound-states in  a medium with GVD
alternating between normal and anomalous dispersion, is discussed. In the
presence of oscillatory gain/loss profile,  we find multiple bound state
structure.

We acknowledge many useful discussions with Prof. G. S. Agarwal.


\begin{references}

\bibitem{GPA}  G. P.  Agrawal, {\it Nonlinear Fiber Optics} (Academic
Press, Inc., San Diego, CA, 2001).

\bibitem{HasegawaB}A. Hasegawa and Y. Kodama, {\it Solitons in Optical
Communications} (Oxford University Press, Oxford, 1995).

\bibitem{Ablowitz} M. J. Ablowitz and Z. H. Musslimani, \PRE {67}{025601(R)}{2003}.

\bibitem{Moores}      J. D.      Moores,     \OL     {21}{555}{1996}.

\bibitem{Kruglov}  V.I .  Kruglov, A. C. Peacock and J. D. Harvey, \PRL
{90}                         {113902}                        {2003}.

\bibitem{PKP}  T. S.  Raju,  P. K.  Panigrahi  and  K. Porsezian, \PRE
{71}{026608}{2005}.


\bibitem{Mollenauer}  L. F. Mollenauer, R. H. Stolen, J. P. Gordon, and
W. J. Tomlinson, \OL {8}{289}{1983}.


\bibitem{Dianov} E. M. Dianov, P. V. Mamyshev, A. M. Prokhorov, and S. V.
Chernikov, \OL {14}{1008}{1989}.

\bibitem{Lisak}  M. L. Quiroga-Teixeiro, D. Anderson, P. A. Andrekson, A.
Bernson and  M.  Lisak,  J. Opt. Soc. Am. B {\bf 13}, 687 (1996).


\bibitem{Malomed1} R. Driben and B. A. Malomed, \PLA {301}{19}{2002}.


\bibitem{Serkin}   V. N.   Serkin  and  A.  Hasegawa,  IEEE  J.  Sel.
Top. Quantum Electron. {\bf 8}, 1 (2002).

\bibitem{Brustev} S. P. Brustev, A. V. Mikhailov, and V. E. Zakharov, Theor. Math. Phys.
{\bf 70}, 227 (1987).

\bibitem{SerkinPRL} V. N. Serkin, A. Hasegawa, and T. L. Belyaeva,
\PRL {92}{199401}{2004}.


\bibitem{Kumar} S. Kumar and A. Hasegawa, \OL
{22}{372}{1997}.

\bibitem{Gabitov} I. Gabitov and S. K. Turitsyn, \JETP
{63}{863}{1996}; I. R. Gabitov and S. K. Turitsyn, \OL {21}{327}{1996}.

\bibitem{Atre}  R.  Atre,  P. K.  Panigrahi  and  G. S.  Agarwal, \PRE
{73}{056611}{2006}.

\bibitem{Hancock} H. Hancock, {\it Theory of Elliptic Functions} (Dover, New York, 1958);
M.   Abramowitz   and  I.  Stegun,  {\it  Handbook  of  Mathematical
Functions}    (NBS,    US   Government   Printing   Office,   1964).

\bibitem{Belanger} P. A. Belanger and N. Belanger, \OC {117}{56}{1996}.


\bibitem{Kruglov1} V. I. Kruglov, A. C. Peacock and J. D. Harvey, \PRE
{71}{056619}{2005}; T. S. Raju, P. K. Panigrahi and K. Porsezian, \PRE
{72}{046612}{2005}; V. I. Kruglov and J. D. Harvey, \JOSAB
{23}{2541}{2006}.

\end{references}
\end{document}